\documentclass[english, letterpaper, 10pt]{article}

\usepackage[T1]{fontenc}
\usepackage{textcomp}

\usepackage{graphicx}
\usepackage{amssymb,amsmath}
\usepackage{fixltx2e}
\usepackage{fancyhdr}
\usepackage{algorithm2e}
\usepackage{url}
\usepackage{makecell}
\usepackage{subfigure}
\usepackage{caption}

\usepackage[numbers,sort&compress]{natbib}

\usepackage{dsfont}
\usepackage[top=1in, bottom=1.25in, left=0.875in, right=0.875in]{geometry}
\usepackage{bm}

\usepackage{mathptmx}   % turns on Times New Roman font
\usepackage{titlesec}
\usepackage{titling}

\usepackage{booktabs}
\usepackage{graphicx}
\usepackage{multirow}

\usepackage[english]{babel}
\usepackage[utf8]{inputenc}
\usepackage{amsmath}
\usepackage{amsfonts}
\usepackage{graphicx}
\usepackage{enumitem}
\usepackage[colorinlistoftodos]{todonotes}
\usepackage{abstract}

\usepackage{indentfirst}%Indent in first paragraph

\usepackage{times}%switch to Times New Roman font

\usepackage{titlesec}% formatting of section title
\titleformat{\section}{\bfseries\centering\fontsize{11pt}{13pt}\selectfont}{\thesection.}{4pt}{\uppercase}
\titleformat{\subsection}{\bfseries\fontsize{11pt}{13pt}\selectfont}{\thesubsection}{6pt}{}
\titlespacing*{\section}{0pt}{10pt plus 8pt}{4pt}
\titlespacing*{\subsection}{0pt}{6pt}{3pt}

\usepackage{titling}% formatting of article title
\pretitle{\vspace{-15mm}\begin{center}\bfseries\LARGE}
\posttitle{\end{center}}
\preauthor{\begin{center}\large}
\postauthor{\end{center}\vspace{-15mm}}

\usepackage{etoolbox}
\apptocmd{\thebibliography}{\setlength{\itemsep}{-2pt}}{}{}

\newlength{\myitemsep}
\setlist[itemize]{itemsep=-1pt, topsep=2pt}
\setlist[enumerate]{itemsep=-1pt, topsep=1pt}

\newcommand{\Lcal}{\mathcal{L}}

\newcommand{\E}[2][]{%
	\ifthenelse{\equal{#2}{}}{\mathbb{E}\left[\,#1\,\right]}{\mathbb{E}_{#1}\left[\,#2\,\right]}%
}

\title{Steganographic Generative Adversarial Networks}

\author{Denis Volkhonskiy$^{1}$, Ivan Nazarov$^{1}$, Evgeny Burnaev$^{1}$
\\~\\
$^{1}$Skolkovo Institute of Science and Technology\\
\textit{Nobel street, 3, Moscow, Moskovskaya oblast’, Russia}\\
e-mail: e.burnaev@skoltech.ru
}

\date{}

\begin{document}
\pagenumbering{gobble}
\maketitle

\renewcommand{\abstractname}{\vspace{0pt}\fontsize{11pt}{13pt}\selectfont \uppercase{Abstract}\vspace{-4pt}}
\begin{abstract}
\normalsize
Steganography is collection of methods to hide secret information (``payload'') within non-secret information ``container''). Its counterpart, Steganalysis, is the practice of determining if a message contains a hidden payload, and recovering it if possible. Presence of hidden payloads is typically detected by a binary classifier. In the present study, we propose a new model for generating image-like containers based on Deep Convolutional Generative Adversarial Networks (DCGAN). This approach allows to generate more setganalysis-secure message embedding using standard steganography algorithms. Experiment results demonstrate that the new model successfully deceives the steganography analyzer, and for this reason, can be used in steganographic applications.
\vspace{0.1mm}
\end{abstract}

\section{Introduction}

Recent years have seen significant advances in estimation methods
and application of deep generative models. There are two major general frameworks for
learning deep generative models: Variational Autoencoders (VAEs),~\cite{kingmawelling2013},
and Generative Adversarial Networks (GANs),~\cite{goodfellowetal2014}. The recent work
of Hu et al.~\cite{huetal2017} develops a unifying framework, which establishes strong
connections of these approaches to Adversarial Domain Adaptation (ADA),~\cite{ganinetal2015}.

GANs have achieved impressive results in semi-supervised learning,~\cite{kingmaetal2014},
and image-to-image translation,~\cite{isolaetal2016}. In~\cite{goodfellowetal2014} the
success of GANs framework was illustrated on the problem of image generation. A more
recent paper~\cite{radfrodetal2015} proposed as set of constraints on the architecture
of convolutional GANs and showed that thus restricted deep convolutional GANs (DCGANs)
are capable of learning a hierarchy of representations from object parts to scenes,
which are sufficiently robust to transfer across domains.

In this study we apply the DCGAN framework to the domain of steganography, i.e. practical
approaches to concealing information (payload) within another piece of information
(stego-container). In particular, we train a generator, images produced by which are
less susceptible to steganographic analysis compared to the original images used as
stego-containers. At the same time, we require that the induced distribution of the
synthetic images approximate well the distribution of the real images in the dataset.

Thus we train a generative model for image stego-containers by confronting it with
two deep convolutional adversaries: a discriminator network, which regularizes the
output to look like samples from the real dataset, and a steganographic analyzer,
which aims at detecting if an image conceals a hidden message. The presence of two
regularizers in the generator's objective resembles the recently proposed multi-target
GAN framework ~\cite{durugkaretal2016}.

\section{Steganography} % (fold)
\label{sec:steganography}

Steganography is a set of algorithms for concealing information in inconspicuous-%
looking communication and a collection of methods to detect and recover the hidden
message from suspicious media (Steganalysis). In steganography the information to be
hidden, the {\it payload}, is embedded by an algorithm inside a cover medium, the
{\it stego-container}. The key drawback is that steganography offers security through
obscurity: an embedded message is sent in the hope that a third party won't detect
or discover it. This makes pure steganography impractical without cryptography, which
deals with secure communication over an insecure channel: the message is scrambled
and authenticated with some keyed algorithm before being concealed in a {\it cover
medium},~\cite{maitietal2011, cheddadetal2010}. In this respect steganography serves
as a layer of weak security by adding a cyphertext detection and extraction step:
encrypted data has much higher entropy than the regular data. Besides information
protection and covert communication, steganography is useful for watermarking in
digital rights management and user identification.

The simplest and most popular algorithm of unkeyed stego-embedding is called the Least
Significant Bit (LSB) matching. The main idea is to take the binary representation
of a secret message, pad it, and store it in a stego-container by overwriting the
LSB of each byte within. The cover-media used for the LSB embedding must be resilient
with respect to bit-level augmentation. In the case of images the least significant
bits of each colour channel of each pixel in the given image are used to hide the payload.

The perturbations introduced by the LSB algorithm do not preserve marginal or joint
colour statistics, which despite being imperceptible to a human observer, simplify
detection of the hidden payload with machine learning or statistical models. A modification
of this method, which addresses this issue to some extent, is the so-called $\pm1$-%
embedding: each bit of the message is hidden by randomly adding or subtracting $1$
from a pixel's colour channel so that the last bit matches it.

More sophisticated steganographic schemes modify the digital media adaptively. The
key idea is to constrain the embedding to regions of high local entropy, e.g. complex
textures or noise. Each pixel is assigned an embedding cost and the embedding locations
are picked in such a way as to minimize the distortion function
\begin{equation*}
  D(I, \hat{I}) = \sum_{i,j} \rho_{ij}(I, \hat{I}_{ij}) \,,
\end{equation*}
where $I$ is the cover image, $\hat{I}$ is the stego-embedding, and $\rho_{ij}(I, 
\hat{I}_{ij})$ is the bounded cost of altering a pixel $(i, j)$ in the cover-image
$I$. The embedding itself is performed by coding methods such as Syndrome-Trellis
Codes (STC)~\cite{filleretal2011}, which are essentially binary linear convolutional
codes represented by parity-check matrix. The state-of-the-art content-adaptive stego-%
embedding algorithms include HUGO~\cite{pevnyetal2010b}, which computes the embedding
costs based on Subtractive Pixel Adjacency Matrix (SPAM) features~\cite{pevnyetal2010a};
WOW~\cite{holubfridrich2010} and S-UNIWARD~\cite{holubetal2014}, which use directional
wavelet filters to weigh and pick regions with high entropy, but implement different
embedding cost functional.

\subsection{Steganalysis}

The simplest approach to steganalysis is based on special feature extractors, e.g.
SPAM~\cite{pevnyetal2010a}, SRM~\cite{fridrichkodovsky2012}, combined with traditional
machine learning models, such as support vector classifiers, decision trees, classifier
ensembles, et c. With the recent overwhelming success of deep learning, specifically
in the image classification and generation domain, newer approaches based on deep
Convolutional Neural Networks (deep CNN) are gaining popularity. For example, in~%
\cite{qianetal2015} it is shown that deep CNN with Gaussian activation functions
achieve competitive performance with hand-crafted features, and in~\cite{pibreetal2015}
it is demonstrated that even shallow CNN are able to outperform the usual ML based
stego-analysis techniques in terms of the detection accuracy.

In this paper we consider steganographic embedding of random bit messages into specifically
crafted images using the $\pm1$-embedding algorithm. The security of the stego-containers
is tested against a class of deep convolutional neural network stego-analyzers, which
try to distinguish images with hidden data from the empty ones.

\subsection{Problem Statement}

\begin{figure}[h!]
  \centering
    \includegraphics[width=15cm]{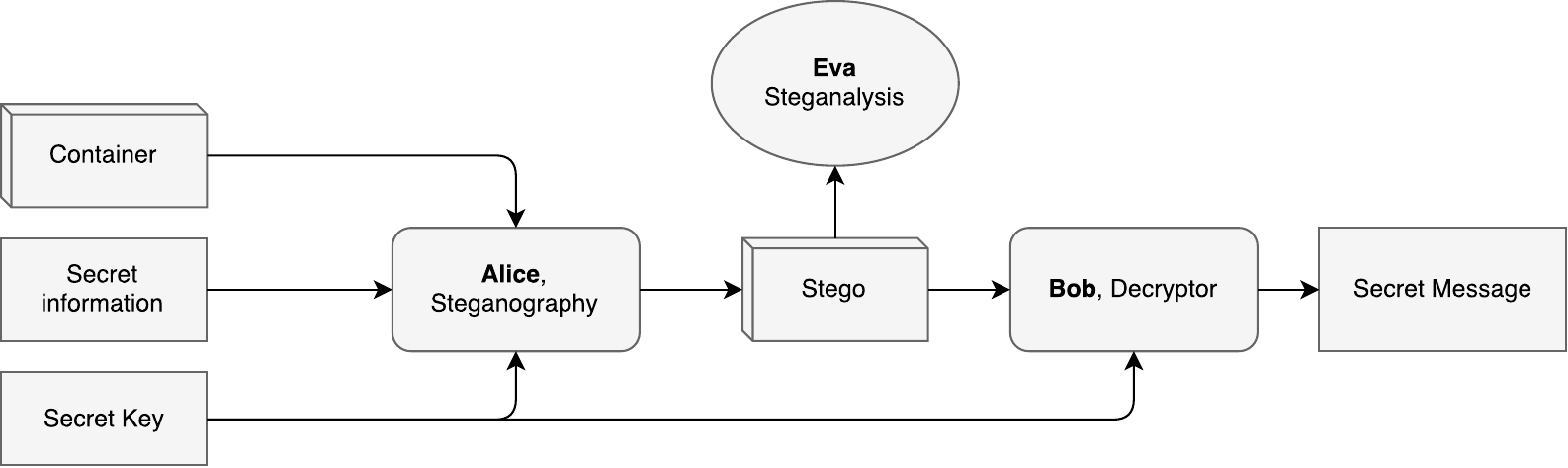}
    \caption{Complete scheme of steganography and steganalysis}
    \label{fig:stego_scheme}
\end{figure}

 The total scheme of steganography and steganalysis is presented at Fig. \ref{fig:stego_scheme}:
\begin{itemize}
\item Usually all images are attacked by a stegoanalyser (Eve);
\item Alice (Steganography algorithm) tries to deceive Eve;
\end{itemize}

The disadvantage of the standard steganography approach is inadaptability of the containers and algorithms for Eve. By this we mean that containers don't adopt (and even know) for type of Eve. The goal for this work is to create adaptive containers generator and steganography new method.

\subsection{Tasks for the research}

We would like to obtain adaptability of the containers to the given Steganalysis in order to deceive it. We set the following tasks for the current work:

\begin{enumerate}
\item Adaptive containers generation.
\begin{itemize}
\item Create a model for image containers generation, that can be used with any Steganography algorithm;
\item Using of these generated containers should deceive Eve (steganalysis);
\item Containers should be adaptive to any type on Eve.
\end{itemize}

 \item New Steganography method:
\begin{itemize}
\item  Create a model for adaptive generation of images with hidden information inside;
\item Test the quality of encryption-decryption on MNIST and CIFAR-10 datasets
\end{itemize}
\end{enumerate}

The difference between these two tasks is the following. 

In the first model, we would like to build a generator of empty containers (images). This images could be used with any Steganography algorithm.

In the second model, we would like to generate not only empty images, but to encode the information to them for further extraction. In other words, we would like to obtain an analogy of visual markers (such as QR-codes).

\section{Generative Adversarial Networks}
Generative Adversarial Networks (GANs) training,~\cite{goodfellowetal2014}, is a
powerful framework for estimating generative models in unsupervised learning setting
by way of a two-player minimax game. The generator player attempts to mimic the true
data distribution $p_\mathtt{data}(x)$ by learning a transformation function $z \mapsto
G_\theta(z)$ of random input values $z$, drawn from a tractable distribution $p_z$.
The generator receives feedback from the discriminator $D_\phi$, which strives to
distinguish synthetic, ``fake'' samples $x=G(z)$, $z \sim p_z$, from genuine, ``real''
samples $x \sim p_\mathtt{data}$.

In the original formulation,~\cite{goodfellowetal2014}, the learning process of the
generator ($G$) and the discriminator ($D$) consists of searching for a saddle point
solution of the following optimization problem
\begin{equation} \label{eq:adv_net_function}
  \begin{aligned}
    & \underset{\theta}{\min}\,\underset{\phi}{\max}
      & & \Lcal(\theta, \phi) =
          \E[x\sim p_\mathtt{data}]{\log D(x; \phi)} 
     + \E[z\sim p_z]{\log(1 - D(G(z; \theta); \phi))}
    \,,
  \end{aligned}
\end{equation}
where $D(x; \phi)$ is the probability output by the player $D$ that $x$ is a real
sample rather then synthetic, and $G(z; \theta)$ is the generated sample. In a typical
application the ground truth data distribution is provided implicitly through its
finite sample approximation on the dataset $\mathcal{D}=(x_i)_{i=1}^m$. Furthermore,
the expectations in the are approximated by sample averages over randomly drawn mini-%
batches of $(G(z_i;\theta))_{i=1}^B$ for $(z_i)_{i=1}^B\sim p_z$ i.i.d. and
$(x_i)_{i=1}^B$ sampled without replacement from the training set.

Despite many advantages, such as generation with a single forward pass and asymptotically
consistent data distribution estimation, the major disadvantage of GANs is that training
them requires finding a Nash (best response) equilibrium,~\cite{goodfellow2016}. Furthermore,
since for deep networks $G(\cdot;\theta)$ and $D(\cdot;\phi)$ the objective is non-convex
w.r.t. the parameters $\phi$ and $\theta$, the order of $\min$ and $\max$ in \eqref{eq:adv_net_function}
matters. Therefore~\cite{goodfellowetal2014} propose to solve the problem by iteratively
alternating between SG maximization and minimization steps, but giving optimizational
advantage to the discriminator player. The idea is to make several gradient ascent steps
on $\Lcal(\theta, \phi)$ w.r.t. $\phi$, before a single gradient descent step w.r.t. $\theta$.

By giving advantage to the discriminator, the proposed approach attempts to approximate
what in fact is the generator's true objective: $\theta \mapsto \max_\phi \Lcal(\theta,
\phi)$. However, since in practice may GANs are trained with alternating single-step
updates,~\cite{nowozinetal2016} propose and justify a simpler joint single-step gradient
method: the SGD update moves along the joint direction $(\nabla_\theta \Lcal, -\nabla_\phi \Lcal)$
obtained via a single back-propagation step. However, there is no consensus as to what
the best training scheme for solving~\eqref{eq:adv_net_function} is,~\cite{goodfellow2016}.

GAN training is also complicated by the different regimes the networks undergo in
the process. For instance, during early stage of training the discriminator is prone
to becoming excessively powerful, which makes the last term of \eqref{eq:adv_net_function}
provide weak feedback to the generator. In~\cite{goodfellowetal2014} authors suggest
to use $\Lcal_\mathtt{Gen}(\theta; \phi) = - \E[z\sim p_z]{\log D(G(z; \theta); \phi)}$
as the minimization objective of the generator, and to set $- \Lcal(\theta, \phi)$ as
the discriminator's loss, $\Lcal_\mathtt{Dis}(\phi; \theta)$, to be {\it minimized}. In
spite of changing the loss and making the game no longer zero-sum, this heuristic leads
to the same saddle point as demonstrated in~\cite{nowozinetal2016}.

\begin{figure}[h!]
	\centering
	\includegraphics[width=8cm]{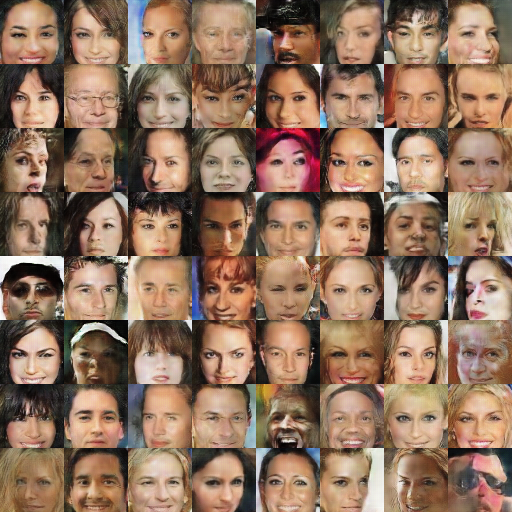}
	\caption{Sample synthetic images generated by DCGAN}
	\label{fig:DCGAN_imgs}
\end{figure}

In fig.~\ref{fig:DCGAN_imgs}
we depict a sample of synthetic images of a freshly trained DCGAN on the Celebrities dataset
\cite{liu2015faceattributes}. The images indeed look realistic, albeit with occasional artifacts.

\section{Steganographic Generative Adversarial Networks}
\label{sec:sgan_model}
\subsection{Model description}
Let $\mathcal{I} = {[-1, 1]}^{H\times W \times 3}$ be the space of images with dimensions
$H\times W$ and RGB channel saturation values between $-1$ and $1$. Let $p_z$ be a
uniform distribution on $\mathcal{Z} = {[-1, 1]}^{d_Z}$, and $p_\mathtt{data}$ be the
distribution of reference images on a subset of $\mathcal{I}$. Finally, the set of
messages $\mathcal{M}$ is given by $\{0, 1\}^{d_M}$, $d_M \leq H W$, and by $S_m: %
\mathcal{I}\mapsto \mathcal{I}$ we denote the LSB embedding of the message $m\in %
\mathcal{M}$ in the cover image $x \in \mathcal{I}$.

\begin{figure}[h!]
  \centering
    \includegraphics[width=10cm]{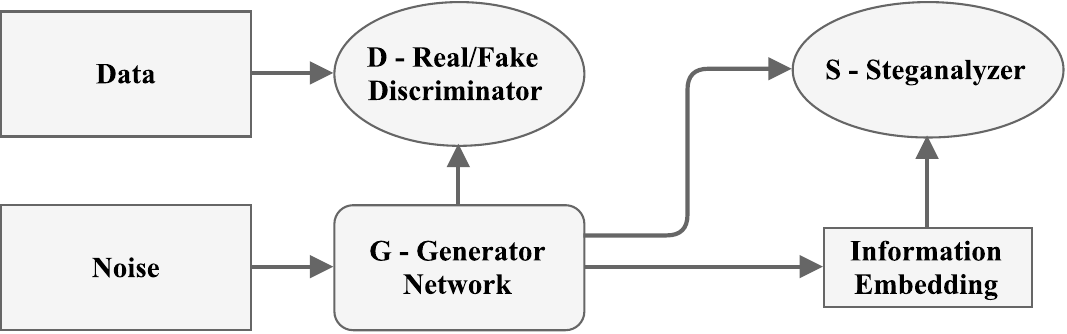}
    \caption{SGAN information flow diagram} \label{fig:SGAN_scheme}
\end{figure}

We introduce Steganographic Generative Adversarial Networks model (SGAN), which is a
zero-sum game between two players: a generator network $G: \mathcal{Z} \mapsto \mathcal{I}$,
which tries to mimic $p_\mathtt{data}$, and an adversary consisting of two parts
\begin{itemize}
  \item a discriminator network $D: \mathcal{I} \mapsto [0,1]$, which distinguishes
  synthetic images $x = G(z)$, $z\sim p_z$, from real $x\sim p_\mathtt{data}$;
  \item a steganalyzer network $A: \mathcal{I} \mapsto [0,1]$, which tries to separate
  cover images $S_m(x)$ with payload $m\in \mathcal{M}$, from empty images $x$, $x\sim Q$,
  for some image distribution $Q$ on $\mathcal{I}$.
\end{itemize}
The value function of the game is fig.~\ref{fig:SGAN_scheme}
\begin{equation} \label{eq:sgan_loss}
  \Lcal(\theta, \phi, \psi) =
    \alpha \Lcal_\mathtt{Dis}(\phi; \theta) +
    (1 - \alpha) \Lcal_\mathtt{San}(\psi; \theta) 
      \to \min_\theta \max_{\phi,\psi} \,,
\end{equation}
\begin{equation} \label{eq:sgan_loss_dis}
  \Lcal_\mathtt{Dis}(\phi; \theta) = 
      \E[x\sim p_\mathtt{data}]{\log D(x; \phi)} + \\
      \E[z\sim p_z]{\log(1 - D(G(z; \theta); \phi))} \,,
\end{equation}
\begin{equation} \label{eq:sgan_loss_san}
  \Lcal_\mathtt{San}(\psi; \theta) = 
      \E[z\sim p_z]{ \E[m]{\log A(S_m(G(z; \theta)); \psi)} } + \\
      \E[z\sim p_z]{\log(1 - A(G(z; \theta); \psi))} \,.
\end{equation}
Here the discriminator and the steganalyzer {\it maximize} the likelihood $\Lcal_%
\mathtt{Dis}$ and $\Lcal_\mathtt{San}$, respectively, while the generator {\it %
minimizes} the convex combination of likelihoods of $D$ and $A$ in~\eqref{eq:sgan_loss}.
The mixing parameter $\alpha \in (0, 1)$ controls the trade-off between the importance
of realism of generated images and their quality as containers against the steganalysis.
Analysis of preliminary experimental results showed that for $\alpha \leq 0.5$ the
generated fails to approximate the distribution of the reference images.

This model resembles the recently proposed multi-target GANs framework,~\cite{durugkaretal2016},
which pits the generator against multiple discriminators: a boosted ensemble, a mean
aggregated combination of discriminators, or discriminators with adaptively adjustable
power. The main idea of the paper naturally suggests that additional discriminators
can be used for regularization, adaptation of the generator's output to other domains,
or endowing the synthetic samples with certain required properties.

The we propose to train the model with the joint simultaneous gradient update scheme
proposed in~\cite{nowozinetal2016}: jointly backpropagate through the networks and
update
\begin{itemize}
  \item for $D$ with $\phi \leftarrow \phi + \gamma_D \nabla_\phi \Lcal_\mathtt{Dis}$;
  \item for $A$ with $\psi \leftarrow \psi + \gamma_A \nabla_\psi \Lcal_\mathtt{San}$;
  \item for $G$ with $\theta \leftarrow \theta - \gamma_G \nabla_\theta \Lcal$, where
  $\Lcal$ is as in~\eqref{eq:sgan_loss}.
\end{itemize}
The expectations are substituted by the empirical averages over the joint mini-batch
of images $x\sim p_\mathtt{data}$, noise $z\sim p_z$, and messages $m \sim \{0, 1\}^{d_M}$.

It is expected that the SGAN game would, for suitable class of deep CNN and an appropriate
training schedule, yield an equilibrium in which the generator produces realistic images
capable of concealing messages embedded by LSB against a deep convolutional steganalyzer.
As is in the case of the original formulation~\cite{goodfellowetal2014}, if the players
were not confined to the class of deep convolutional networks, the optimal steganalyzer
would have been given by the ratio estimator between the distribution of the images $q$,
induced by $G(z;\theta)$ for $z\sim p_z$, and the distribution $q_s$, implicitly defined
as the distribution of $S_m(x)$ over $x\sim q$ and $m\sim \{0,1\}^{d_M}$. We expect that
the optimal generator would induce a distribution, the variates from which would have
uniformly random least significant bit in the colour data of each pixel, since for a
random $m$ the stego-embedding $S_m(x)$ alters $x$ only on the scale of $2^{-7}$, and
the result is, essentially, a random bit.

% subsection description (end)

\subsection{Challenges} % (fold)
\label{sub:challenges}

One of the challenges in the proposed model is the fact that any stego-embedding algorithm
introduces a distortion to the cover medium, which is generally a non-differentiable
perturbation as a function of the data in the medium. For instance, the LSB matching
modifies the colour channel data of the pixels independently, and thus can be represented
as a residual-like transformation of each colour channel value: $S_m(x) = x + \delta_m(x)$,
for any message bit $m \in \{0,1\}$ and $x \in [-1, +1]$, where $\delta_m(x) \in \{0,%
\pm\epsilon\}$, $\epsilon = 2^{-7}$, is the distortion. The $\delta_m(x)$ is a random
function given by
\begin{equation} \label{eq:lsb_distortion}
  % S_m(x) = x + \delta_m(x)\,,
  \delta_m(x) = \xi \, \mathbf{p}_m\bigl((x + 1) \epsilon^{-1} \bigr) \,,
  % \mathbf{p}_m(z) = \mathbf{1}_{\{\exists k\in \mathbb{Z} \,:\, z - m \in [2k, 2k+1)\}} \,,
  % \mathbf{p}_m(z) = \sum_{k \in \mathbb{Z}} \mathbf{1}_{[2k, 2k+1)}(z - m) \,.
\end{equation}
where $\mathbf{p}_m(z)$ is $0$ when $\exists k \in \mathbb{Z} \,:\, z - m \in [2k, 2k+1)$,
i.e. the LSB of the integer part of $z$ matches $m$, and $1$ otherwise. The value $\xi$
is $+\epsilon$ if $m=1$ and $x \in [-1, -1+\epsilon)$, and $-\epsilon$ if $m=0$ and
$x \in (+1-\epsilon, +1]$, but otherwise an independent random variable $\pm\epsilon$,
which is the addition~/~subtraction mask in the LSB algorithm.

The key observation is that for fixed $m$ and $\xi$ this perturbation is fully determined
by $\mathbf{p}_m(z)$, which is {\it constant} on the intervals $(k, k + 1)$ with $k\in%
\{0 \ldots 255\}$. While this function is non-differentiable at finitely many points,
where the $0-1$ switching occurs, everywhere else in $(0, 256)$ it is constant and
has zero derivative w.r.t. $z$. Thus for fixed $m$ and $\xi$ the distortion $\delta_m(x)$
has derivative zero almost for every $x \in (-1, 1)$. In light of this heuristic argument,
the following procedure for stego-embedding can be used while training: during the
forward pass the embedding $x\mapsto S_m(x)$ is exact, but during the back propagation
the embedding response is approximate by an identity function $S_m(x) \approx x$.
% We trade the non-differentiability issues for a possible risk of a weaker steganalyzer

The main drawback of this ``linear'' approximation is that it provides essentially no
gradient feedback to the generator and acts as if nothing was embedded. Thus, although
being almost correct it is ill suited for the purpose of learning stego-secure cover
entities by design. Therefore we propose another approach: for a fixed message $m\in %
\{0, 1\}$ and noise $\xi\in \{\pm 1\}$ approximate the LSB embedding layer by a differentiable
transformation. To this end we substitute the mismatch indicator in~\eqref{eq:lsb_distortion}
by a sine waveform with nonlinear gain at extremes of its range:
\begin{equation}\label{eq:lsb_distortion_approx}
  \mathbf{p}_m(z)
    \approx \mathbf{s}_\beta(z; m)
    = \sigma\bigl(\beta \sin{(m - z) \pi} \bigr) \,,
\end{equation}
where $\beta > 0$ determines the fidelity of the approximation and $\sigma(a)$ is
the {\bf Sigmoid} function, $a\mapsto {(1+e^{-a})}^{-1}$. The derivative of the {\it%
Sigmoid} function is $a\mapsto \sigma(a)(1-\sigma(a))$, and therefore this approximation
is not computationally too demanding and provides accurate gradient feedback at the
jump points of $\mathbf{p}_m(z)$.

\subsection{Training process}

Stochastic mini-batch Gradient descent update rules for components of SGAN are listed below:
\begin{itemize}
	\item for $D$ the rule is $\theta_{D} \leftarrow \theta_{D} +  \gamma_{D} \nabla_{G} L$
    with
	\begin{equation*}
      \nabla_{G} L = \frac{\partial }{\partial \theta_{D}} \Bigl\{
	      \E[x\sim p_{data}(x)]{\log D(x,\theta_{D})}
    	  + \E[z\sim p_{noise}(z)]{\log(1 - D(G(z,\theta_{G}),\theta_{D}))}
      \Bigr\} \,;
    \end{equation*}  
	\item for $S$ (it is updated similarly to $D$):
    $\theta_{S} \leftarrow \theta_{S} +  \gamma_{S} \nabla_{S} L$
    where
    \begin{equation*}
	    \nabla_{S} L = \frac{\partial}{\partial \theta_{S}}
        	\E[z\sim p_{noise}(z)]{\log S(Stego(G(z,\theta_G)),\theta_{S})
            				 + \log(1 - S(G(z,\theta_G),\theta_{S}))}\,;
    \end{equation*} 
	\item for the generator $G$: $\theta_{G} \leftarrow \theta_{G} -  \gamma_{G} \nabla_{G} L$
    with $\nabla_{G} L$ given by
    \begin{align*}
      	\nabla_{G} L
        	&= \frac{\partial }{\partial \theta_{G}}
            	\alpha \E[z\sim p_{noise}(z)]{\log (1-D(G(z,\theta_{G}),\theta_{D}))} 
			+ \frac{\partial }{\partial \theta_{G}}
            	(1-\alpha) \E[z\sim p_{noise}(z)]{\log(S(Stego(G(z,\theta_{G}),\theta_{S})))}\\
			&+ \frac{\partial }{\partial \theta_{G}}
            	(1-\alpha)  \E[z\sim p_{noise}(z)]{\log (1-S(G(z,\theta_{G}),\theta_{S}))} \,.
    \end{align*}
\end{itemize}
The main distinction from the GAN model is that we update $G$ in order to maximize not only the error of $D$, but to maximize the error of the linear combination of the classifiers $D$ and $S$.
\section{Steganographic Encryption Generative Adversarial Networks}
\label{sec:segan_model}
\subsection{Model Description}
Steganographic Encryption Generative Adversarial Networks (SEGAN) model was constructed for information encryption/ description purposes. It consists of 
\begin{itemize}
\item Alice --- A generator network: produce realistic images, that contains hidden information.
\begin{itemize}
\item Input: Secret Key (binary), Secret Message (binary), Class for generation (y), noise
\item Output: Image with hidden secret message inside
\end{itemize}
\item Bob --- A decryption network: extract hidden message from the image.
\begin{itemize}
\item Input: Image, Secret Key, Class of the image (y)
\item Output: Secret hidden message
\end{itemize}
\item Discriminator --- A discriminator, that tries to detect whether the image is real or generated.
\begin{itemize}
\item Input: Image, Class of the image (y)
\item Output: 0/1 -- generated/real class
\end{itemize}
\end{itemize}

\begin{figure}[h!]
  \centering
    \includegraphics[width=12cm]{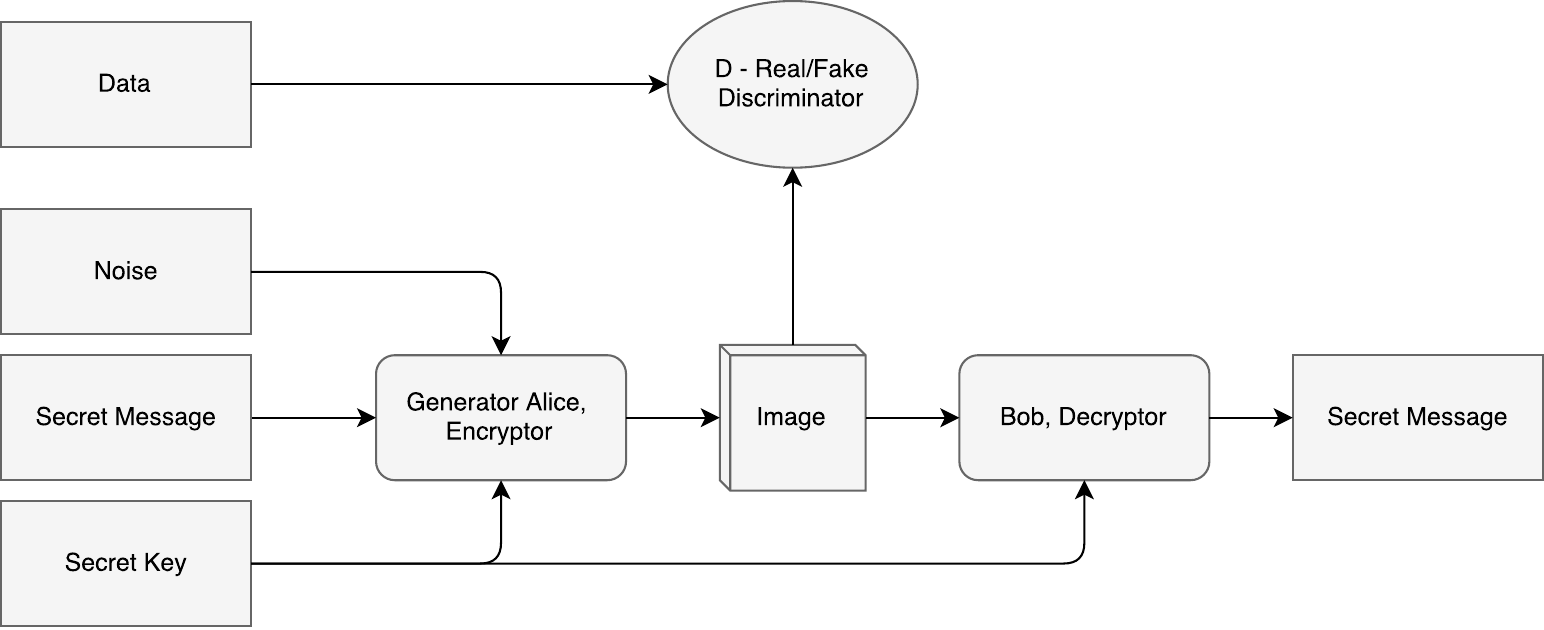}
    \caption{SEGAN information flow diagram} \label{fig:SEGAN_scheme}
\end{figure}

The full scheme is presented in Fig. \ref{fig:SEGAN_scheme}. This model can be considered as a autoencoder with the high-dimensional hidden representation. In such case Discriminator is considered as such kind of regularization.

\subsection{Training process}
The training process can be represented as a usual GAN training process, but with some modification.

Firstly we should define the loss functions for constructing the updating rules: 
\begin{itemize}
\item Alice (as realistic image Generator) loss. Standard GANs generator loss.

\begin{align}\label{eq:alice_gen_loss}
\begin{split}
L_A = &\mathbb{E}_{y} \mathbb{E}_{m \sim p_{message}}\mathbb{E}_{k \sim p_{key}}\mathbb{E}_{z \sim p_{noise}}[\log(1 
- C(\theta_C, (\theta_A, m, k, z, y), y)) \mid y, m, k, z]
\end{split}
\end{align}
\item Alice-Bob encryption loss. Here we compute standard $l2$ loss between the original message and reconstructed message.

\begin{align}\label{eq:alice_bob_encription_loss}
\begin{split}
 L_{AB} =& \mathbb{E}_{y} \mathbb{E}_{m \sim p_{message}}\mathbb{E}_{k \sim p_{key}} \mathbb{E}_{z \sim p_{noise}} 
 \mathbb{E}[(m 
 -  B(\theta_B, A(\theta_A, m, k, z, y), k, y))^2 \mid y, m, k, z] \to \min
\end{split}
\end{align}

\item Discriminator's loss. This is a usual GANs discriminator loss, witch is calculated as an average cross-entropy.

\begin{align}\label{eq:critic_loss}
\begin{split}
 L_C =
   \mathbb{E}_{y}& \E[x \sim p_{data}]{\log(C(\theta_C, x, y)) \mid y} -  \\
  -&\mathbb{E}_{y} \mathbb{E}_{m \sim p_{message}}\mathbb{E}_{k \sim p_{key}}\mathbb{E}_{z \sim p_{noise}} [\log(1 -
   C(\theta_C, A(\theta_A, m, k, z, y), y)) \mid y, m, k, z] \to \min
\end{split}
\end{align}
\end{itemize}

The total SEGAN train procedure is presented in Algorithm \ref{alg:segan_train}. 

\begin{algorithm}[tb]
    \SetKwInOut{Input}{Input}
    \SetKwInOut{Output}{Output}
    \Input{$d_k$ --- The dimension of a binary key\\
    $d_m$ --- The dimension of a binary message,\\ $d_n$ --- The dimension of the input noise}
    \For{epoch in $1\ldots n\_epoch$} {
    \For{minibatch from the data set}{
      Sample minibatch of  $d_n$ noise samples $\{z_1, \ldots, z_{d_n}\}$\\
      Sample minibatch of  $d_m$ noise samples $\{m_1, \ldots, m_{d_m}\}$ \\
      Sample minibatch of  $d_k$ noise samples $\{k_1, \ldots, k_{d_k}\}$ \\
      
      Update $A$ according to loss $L_A$\\
      Update $C$ according to loss $L_C$\\
      \If{$epoch > 1$}{
      Update $A$, $B$ according to loss $L_{AB}$
      }
      }
    }
    \caption{Algorithm of training SEGAN}
    \label{alg:segan_train}
\end{algorithm}

\section{Experiments with Steganographic Generative Adversarial Networks}

\subsection{Steganographic Vectors} % (fold)
\label{sub:steganographic_vectors}

Before conducting extensive numerical experiments on the security of the LSB embedding
in cover images produced by a SGAN trained generator, we run a simpler experiment
as a proof-of-concept. Since the LSB matching embeds each bit of the message into
a pixel independently from the its context, we study numerically the stego-security
properties of SGAN generated $1$-d vectors.

\subsubsection{Validation Protocol} % (fold)
\label{sub:validation_protocol}

In the process of training SGAN for $T$ iterations we obtain a sequence of generators
$(G^t)_{t=1}^T$ where $G^t(\cdot)=G(\cdot; \theta_t)$ and $\theta_t$ are the parameters
of the generator after $t$ minibatch SGD updates. We use the following empirical
validation protocol for the generator after the $t$-th iteration:
\begin{enumerate}
%%%% Holy shit, this is so wrong! The produced sample is not iid: the train-test
%%%%  split on one sample of this sort yields AUC < 0.42, while AUC on a honestly
%%%%  and independently re-generated sample of this kind yieldes AUC ~ 0.51+/-.03
%%% \item get a sample $\mathcal{S}^{t} = (x_i, y_i)_{i=1}^{2 M}$:
%%% \begin{itemize}
%%%   \item synthesize $M$ cover entities $x = G(z)$, $z\sim p_z$;
%%%   \item embed an $m \in\{0,1\}^{d_M}$ in each $x$ with $S_m(\cdot)$;
%%%   \item label $x$ with $y=0$ and $S_m(x)$ with $y=1$.
%%% \end{itemize}
%%%% Preserved this for the posterity to behold.
  \item Draw a sample $\mathcal{S}^t = (x_i, y_i)_{i=1}^M$, i.e. for $i=1,...M$
  \begin{enumerate}
    \item independently draw $y_i \sim \{0,1\}$ and $z_i \sim p_z$;
    \item get a message $m_i \in \mathcal{M} = \{0, 1\}^{d_M}$;
    \item synthesize a cover entity $x^*_i = G^t(z_i)$ and set
     $x_i =
      \begin{cases}
        S_{m_i}(x^*_i), &\text{ if } y_i=1,\\
        x^*_i, &\text{ if } y_i=0.
      \end{cases}$
  \end{enumerate}
  \item Assess the performance of an independent steganalyzer $A^*$ with
  $K$-fold cross validation on $\mathcal{S}^t$.
\end{enumerate}
This sampling procedure ensures that the examples in the stego-sample $\mathcal{S}^t$
are independent and identically distributed.

We also control the diversity of the embedded messages $m_i$ through different message
generation scenarios:
\begin{itemize}
  \item {\bf Fixed}: $m_i = m_0$, with $m_0$ picked once from $\mathcal{M}$;
  \item {\bf Pool ($n$)}: random $m_i$ from $\mathcal{M}_0 \subset \mathcal{M}$,
  $|\mathcal{M}_0| = n$;
  \item {\bf Arbitrary}: random $m_i$ from $\mathcal{M}$.
\end{itemize}
Under the ``Fixed'' and ``Pool ($n$)'' variants $m_0$ and $\mathcal{M}_0$, respectively,
are chosen independently for each sample $\mathcal{S}^t$ in the above outline. The
motivation behind controlling the diversity stems from the idea that by comparing
the optimal performance of $A^*$ under each scenario it is possible to verify if
$S_m(G(z)) \equiv G(z)$ in distribution for independent $z \sim p_z$ and $m\sim %
\mathcal{M}$. In fact, with SGAN it is possible to train a generator, that induces
an distribution, which is invariant under LSB embedding of messages from a specific,
non-uniform distribution of messages $\mathcal{M}$ just by sampling from it during
training.

We use this validation protocol for the experiments in sec.~\ref{sub:steganographic_vectors}.
Howeer, for the experiments with image generation ( sec.~\ref{sub:steganographic_images})
we extend the protocol, since in this case we have a reference distribution $Q$ on
$\mathcal{I}$, which the generator $G$ aims to replicate.

\begin{figure}[!ht]
  \centering
  \includegraphics[width=0.7\linewidth]{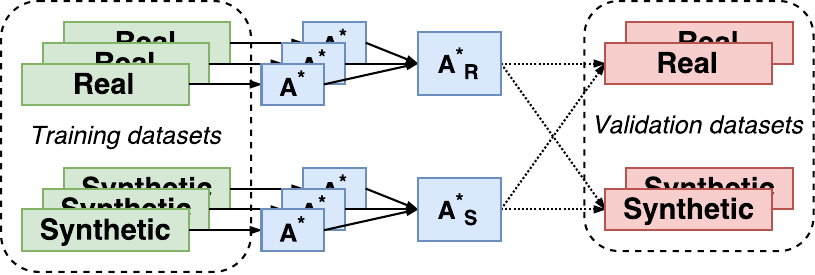}
  \caption{Extended validation protocol for experiments in sec.~%
  \ref{sub:steganographic_vectors}.}
  \label{fig:stegoimage_validation_protocol}
\end{figure}

The extension is illustrated in fig.~\ref{fig:stegoimage_validation_protocol}. Firstly,
several independent real and synthetic training datasets are generated:
\begin{enumerate}
  \item Draw a synthetic stego-sample $\mathcal{S}_\mathtt{S}$ as outlined above;
  \item Generate similarly a real stego-sample $\mathcal{S}_\mathtt{R}$ by sampling $x^*_i$ from $Q$ (sample without replacement if $Q$ is an empirical distribution);
\end{enumerate}
Secondly, on each of these datasets we train an independent steganalyzer $A^*$ and
aggregate them with weighted majority voting to obtain the final pair of steganalyzers
$A^*_\mathtt{S}$ and $A^*_\mathtt{R}$, where the former is based on synthetic training
datasets and the latter -- on real stego-samples. Finally, we independently generate
several real and synthetic datasets as outlined above, and used them for cross-validation
of the steganalyzers: each one is validated on both kinds of test samples to assess
if and how well the learned features transfer across real~/~synthetic domains.

\subsubsection{Details of the Experiment} % (fold)
\label{ssub:details}

The generator $G$ is a $1$-d convolutional neural network (CNN) that expands the input
$z\sim \mathcal{N}_{d_I}(0, I_{d_I})$ from $\mathbb{R}^{d_I}$ into $[-1,+1]^{d_O}$,
where $d_I=4$ and $d_O=16$. The generator uses a series of $1$-d transposed convolution
layers with nonunit strides and the {\bf ReLU} nonliearity, $a\mapsto\max\{a, 0\}$,
to upsample the input noise into a intermediate $32\times d_O$ state matrix, which
is finally passed through a $1$-d convolution layer with {\bf Tanh} nonlinearity,
$a\mapsto \frac{e^a - e^{-a}}{e^a + e^{-a}}$, to ensure ${[-1, +1]}^{d_O}$ output.

The steganalyzer $A$ is also a $1$-d CNN which takes a sample $x\in [-1,+1]^{d_O}$
and feeds the input through a set of $1$-d convolution layers with {\bf Leaky ReLU},
$a \mapsto \max\{\alpha a, a\}$ for $\alpha = \frac1{20}$ to get the $256\times 4$
matrix of intermediate features. The matrix is passed through two fully connected
layers with the {\bf Leaky ReLU} and $512$ units each. The final layer of $A$ outputs
a single logit score, which reflects the degree to which the analyzer is ``confident''
that the sample $x$ contains a message hidden with the LSB embedding. This network
$A$, as well as $G$, has a {\bf Batch Normalization (BN)} layer,~\cite{ioffeszegedy2015},
before the nonlinearity, except for the output layers.

The losses are set-up as in \eqref{eq:sgan_loss} and \eqref{eq:sgan_loss_san}, but
{\bf without} an explicit ``discriminator'' player. We simulate its feedback by
imposing a total variation with $l_2$ loss on the generator's output:
\begin{equation*} \label{eq:sgan_loss_dis_exp1}
  \Lcal_\mathtt{Dis}(\theta) = 
      \Biggl.
        \E[z\sim p_z]{
          \sum_{i=1}^{d_O}
            C_\mathtt{tv} \bigl\lvert x_i - x_{i-1} \bigr\rvert
            + C_\mathtt{l_2} \frac12 x_i^2
        }
      \Biggr\rvert_{\substack{x=G(z;\theta) \\ x_0 = x_{d_O}}}
        \,.
\end{equation*}
The relative weight $\alpha$ in the generator's loss is set to $\alpha = \frac{C_\mathtt{san}}%
{C_\mathtt{san} + C_\mathtt{tv} + C_\mathtt{l_2}}$ for $C_\mathtt{san} = 2$,
$C_\mathtt{tv} = {10}^{-2}$, and $C_\mathtt{l_2} = {10}^{-1}$.

We train SGAN generator for $T=75\cdot{10}^3$ minibatch iterations of $256$ pairs
of noise and message vectors. We set $d_M = d_O$ and thus embed $1$ bit of message
in each element of the cover vector, achieving the flat vector analogue of image's
$1$ bit-per-pixel embedding density.

SGAN networks are trained with Adam,~\cite{kingmaba2014}, with default parameter values
and learning rates $2^{-14}$ and $2^{-18}$ for the steganalyzer ($A$) and the generator
($G$), respectively. We use different learning rates to make the generator adapt slowly
to the steganalyzer's feedback, while allowing the analyzer to learn to detect hidden
data in the generator's output more quickly, \cite{huseletal2017}. The training was
carried out with the strategy proposed in~\cite{goodfellowetal2014}: $4$ updates of
$A$ per $1$ update of $G$. The scale parameter $\beta$ of~\eqref{eq:lsb_distortion_approx}
was set to $8$.

We run the validation protocol with the sample size $M=2^{16}$ and the number of folds
$K$ set to $7$. In this setting we validate two different independent steganalyzers
$A^*$: a {\it Random Forest} classifier ensemble with $101$ trees and a {\it $1$-d CNN}
similar to the network used during SGAN training. The only difference of the CNN $A^*$
is that it yields two values and passes them through the {\bf Softmax} nonlinearity,
$(a,b) \mapsto \bigl(\frac1{1 + e^{b-a}}, \frac1{1 + e^{a-b}}\bigr)$. The convolutional
$A^*$ is trained using Adam SGD updates with learning rate $2^{-14}$ and weight decay
of $2^{-3}$ ($l_2$ parameter regularizer).

To simplify the task of the independent steganalyzer $A^*$ we use hand-crafted features
instead of raw input vectors. These features are derived with the {\it 8-bit reversal}
procedure and are strongly tailored to the LSB embedding. The main idea is that through
``reversal'' of the bit order in each element these features amplify the signal from the
stego-embedding, which is naturally a weak high frequency distortion. Each value $v \in%
[-1,+1]$ in the input vector is transformed as follows:
\begin{itemize}
  \item set $u = (v + 1) \cdot \epsilon$ for $\epsilon = \frac{255}2$ and let
  $\hat{u} = \lfloor u\rfloor$;
  \item convert $\hat{u}$ to an unsigned $8$-bit value;
  % \item reverse its bit order to get $\overleftarrow{u} = $;
  \item produce $\frac1{\epsilon} (\mathtt{rev}(\hat{u}) + (u - \hat{u}))- 1$;
\end{itemize}
where $\mathtt{rev}(\cdot): \{0, 1\}^{8}\mapsto \{0, 1\}^{8}$ reverses the bit order:
$ \label{eq:rev8bit}
  \mathtt{rev}\bigl((b_0, b_1, \cdots, b_7)\bigr)
    = (b_7, b_6, \cdots, b_0) \,,
    % = \mathtt{rev}\biggl(\sum_{k=0}^7 2^k b_k\biggr)
    % = \sum_{k=0}^7 2^k b_{7-k} \,,
$
with the bit at the first position in the sequence being {\it the least significant
bit} (the LSB bit endianness). With these features any change in the least significant
bit would affect the whole value on the $2^7$ scale instead of $2^0$, and make more
prominent the arithmetic effects in the LSB matching due to $\pm1$ addition.

% subsection validation_protocol (end)

\subsubsection{Results} % (fold)
\label{ssub:stegovector_results}
The performance of the Random Forest $A^*$ on the output vectors of the generator
after $T$ iterations is shown in table~\ref{tab:stegovector_kfold_rf_final}, of the
$1$-d CNN -- in Table~\ref{tab:stegovector_kfold_nn_final}. Both tables were obtained
via the validation protocol and the feature preprocessing outlined above. It is worth
noting that in both runs all metrics drop dramatically when the variety of embedded
messages is increased, which lends evidence to the invariance of the induced distribution
under the LSB distortion. Numerical experiments were also conducted on raw features,
but both steganalyzers performed poorly.

\begin{table}[!ht]
  %\normalsize
  %
  \caption{Performance metrics of the Random Forest $A^*$ on the cover vectors
  generated after all iterations of training ($\mu \pm 5\sigma$ scaled by $100$).}
  \centering
  \begin{tabular}{lrrrr}
    % # RF: results for "20170819-121319" feature type "rev(x)"
    \hline
    metric        & ROC-AUC         & F1-score         & Accuracy         & OOB     \\
    type          &                 &                  &                  &         \\
    \hline
    Fixed         &   $100.0 \pm 0.0$ &    $100.0 \pm 0.0$ &    $100.0 \pm 0.0$ & $100.0$ \\\\

    % Pool ($16$)   &   $100.0 \pm 0.0$ &    $100.0 \pm 0.1$ &    $100.0 \pm 0.1$ & $100.0$ \\
    Pool ($64$)   &    $99.9 \pm 0.1$ &     $99.9 \pm 0.2$ &     $99.9 \pm 0.2$ &  $99.9$ \\
    Pool ($256$)  &    $99.7 \pm 0.2$ &     $98.9 \pm 0.6$ &     $98.9 \pm 0.6$ &  $98.8$ \\
    Pool ($1024$) &    $84.7 \pm 3.0$ &     $77.2 \pm 3.2$ &     $76.7 \pm 2.9$ &  $75.2$ \\
    Pool ($4096$) &    $54.8 \pm 2.7$ &     $53.1 \pm 3.8$ &     $53.5 \pm 3.2$ &  $52.4$ \\\\

    Arbitrary     &    $49.6 \pm 3.0$ &     $50.4 \pm 2.7$ &     $49.8 \pm 2.2$ &  $49.9$ \\
    \hline
  \end{tabular}

  \label{tab:stegovector_kfold_rf_final}
\end{table}

\begin{table}[!ht]
  %\scriptsize
  %
  \caption{Performance metrics of the $1$-d CNN $A^*$ on the cover vectors
  generated after all iterations of training ($\mu \pm 5\sigma$ scaled by $100$).}
  \centering
  \begin{tabular}{lrrr}
  % # NN: results for "20170819-121319" feature type "rev(x)"
    \hline
    metric        & ROC-AUC           & F1-score           & Accuracy           \\
    type          &                   &                    &                    \\
    \hline
    Fixed         &   $100.0 \pm 0.0$ &     $99.9 \pm 0.1$ &     $99.9 \pm 0.1$ \\\\

    % Pool ($16$)   &    $99.9 \pm 0.1$ &     $99.1 \pm 0.5$ &     $99.1 \pm 0.5$ \\
    Pool ($64$)   &    $98.9 \pm 0.3$ &     $95.6 \pm 1.3$ &     $95.5 \pm 1.2$ \\
    Pool ($256$)  &    $92.1 \pm 1.7$ &     $84.7 \pm 2.6$ &     $84.5 \pm 2.3$ \\
    Pool ($1024$) &    $70.0 \pm 4.5$ &     $64.9 \pm 5.5$ &     $64.6 \pm 3.7$ \\
    Pool ($4096$) &    $56.6 \pm 2.5$ &     $55.1 \pm 4.8$ &     $54.8 \pm 2.8$ \\\\

    Arbitrary     &    $50.0 \pm 1.9$ &     $49.9 \pm 5.9$ &     $50.2 \pm 2.6$ \\
    \hline
  \end{tabular}
  
  \label{tab:stegovector_kfold_nn_final}
\end{table}

We also experimented with the strength of the feedback to $G$ from the steganalyzer
$A$ in SGAN training by setting $C_\mathtt{tv} = 0$ and changing $C_\mathtt{san}$.
The trained generator appears to be capable of successfully hiding message bits against
the random forest $A^*$ even at the proportion $C_\mathtt{l_2} : C_\mathtt{san}$ as
low as $99 : 1$. This can be attributed to simplicity of the generator's goal in this
experiment. For $C_\mathtt{san} = {10}^{-3}$ and $C_\mathtt{l_2} = 1 - C_\mathtt{san}$,
however, the generator failed to produce LSB-invariant output distribution even in
the ``Arbitrary'' message generation scenario.  % run `20170820-200450`
This implies that the feedback provided by \eqref{eq:sgan_loss_san} is relevant to
the task of generating cover vectors, which make the LSB matching embedding less
susceptible to steganalysis.

\begin{figure}[!ht]
  \centering
  \includegraphics[width=0.8\linewidth]{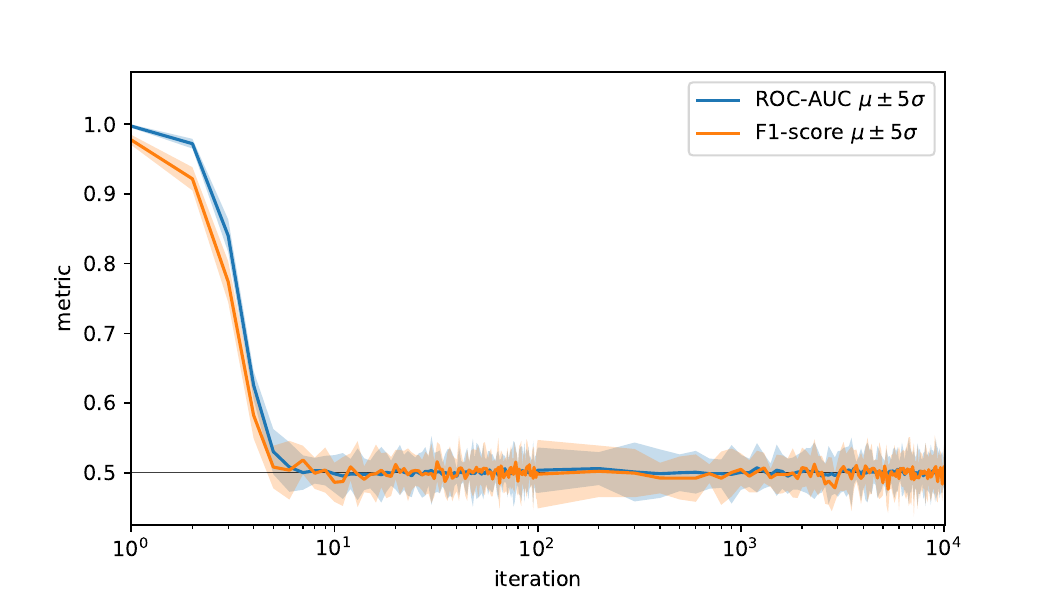}
  \caption{$k$-fold CV performance of $A^*$ (Random Forest) on synthetic images
  produced by the generator at different stages of training.}
  \label{fig:stegovector_kfold_rf}
\end{figure}

\begin{figure}[!ht]
  \centering
  \includegraphics[width=0.8\linewidth]{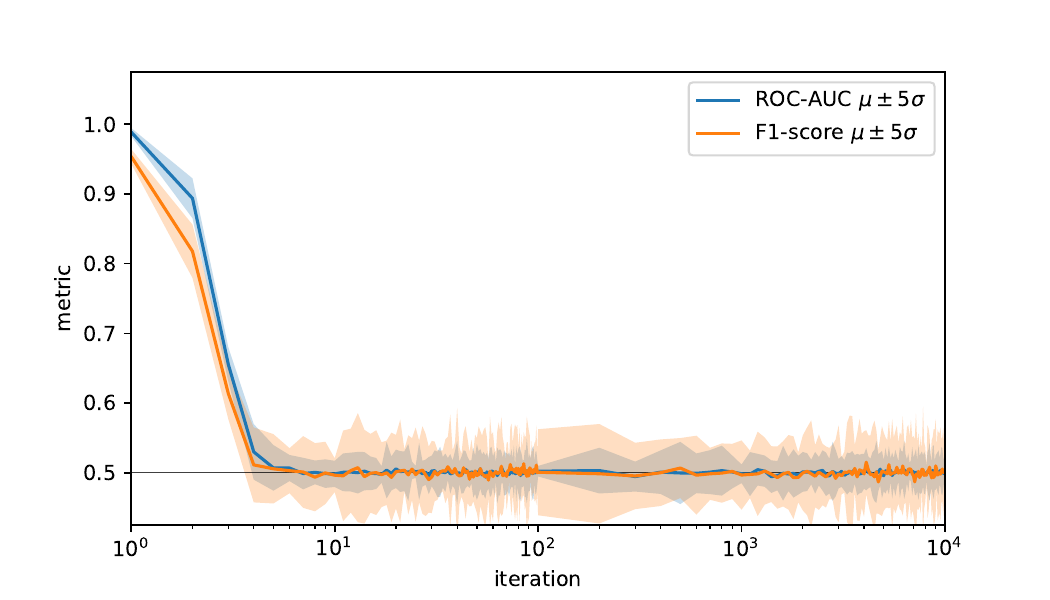}
  \caption{$k$-fold CV performance of $A^*$ ($1$-d CNN) on synthetic images
  generated at different stages during the SGAN trainnig.}
  \label{fig:stegovector_kfold_nn}
\end{figure}

The figures~\ref{fig:stegovector_kfold_rf} and~\ref{fig:stegovector_kfold_nn} show
the performance dynamics of the tested steganalyzers. The generator converges to the
desired output distribution is very quickly: in all SGAN re-runs with moderate to
high values of $C_\mathtt{san}$ the steganalyzer $A^*$ fails to discriminate between
empty and non-empty vectors after at most $20$ of training. Note that in the figures
the proposed validation protocol was performed on each iteration during the first
$100$ SGAN iterations, and only on every $100$-th iteration onward.

The main conclusion from this set of experiments is that the SGAN model~\eqref{eq:sgan_loss}
is useful in generating cover vectors, the unconditional distribution of which is
invariant under the LSB embedding distortion~\eqref{eq:lsb_distortion}. Also the
embedding approximation~\eqref{eq:lsb_distortion_approx} is adequate and provides
relevant gradient feedback. In the next section we experiment with generating realistic
cover images, which make the simple LSB embedding more steganographically secure.
\subsection{Steganographic Images}
\label{sub:steganographic_images}
\subsubsection{Data Description}
In our experiments we use the Celebrities dataset \cite{liu2015faceattributes} that contains $200\,000$ images.
All images were cropped to $64\times 64$ pixels.

\begin{figure}[h!]
	\centering
	\includegraphics[width=9cm]{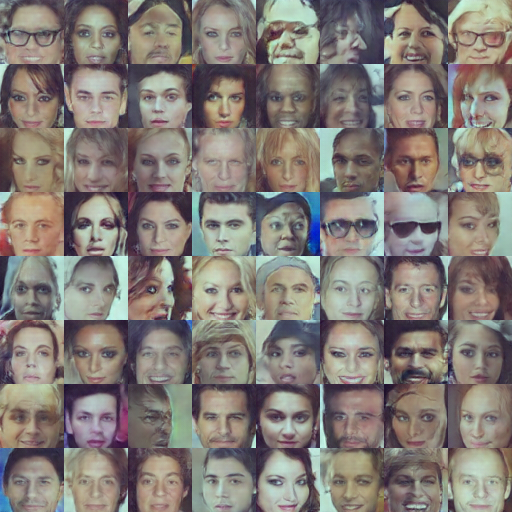}
	\caption{Examples of images, generated by SGAN after training for 8 epochs on
    the Celebrities dataset}
	\label{fig:SGAN_4_epoch}
\end{figure}

For steganalysis purposes we consider $10\%$ of data as a test set. We denote the train set by
$A$, the test set by $B$ and steganography algorithms used for hiding information by $Stego(x)$.
After embedding some secret information we get the train set $A + Stego(A)$ and the test set
$B + Stego(B)$, We end up with $380\,000$ images for steganalysis training and $20\,000$ for
testing. For training the SGAN model we used all $200\,000$ cropped images. After $8$ epochs
of training our SGAN produces images displayed in fig.~\ref{fig:SGAN_4_epoch}.

For information embedding we use the $\pm1$-embedding algorithm with a payload size equal to
$0.4$ bits per pixel for only one channel out of three. As a text for embedding we use randomly
selected excerpts from some article from The New York Times.

\subsubsection{Experimental Setup}
In this section we describe the SGAN model structure. By \emph{C2D-BN-LR} we denote the
following structural block of a convolutional neural network: Conv2d $\rightarrow$ Batch
Normalization $\rightarrow$ Leaky ReLU.

\noindent The Steganalyser network $S$ and the Image Discriminator network have similar
structure: four \emph{C2D-BN-LR} layers, then a fully connected layer ($1$ neuron) $\rightarrow$
Sigmoid function is used to compute an output. The Image generator network $G$ is (in order)
a fully-connected layer ($8192$ neurons), four \emph{C2D-BN-LR} with Fractional-Strided
convolution, then the Hyperbolic tangent function layer is used to compute normalised output.

The SGAN model is trained to solve \eqref{eq:sgan_loss} using the Adam optimization algorithm \cite{kingma2014adam} with the learning rate $2^{-4}$ and update parameters $\beta_1 = 0.5$
and $\beta_2 = 0.999$. For each mini-batch of images we update weights of $D$ and $S$ once,
then we update weights of $G$ twice.

In the following experiments, in addition to the steganalyser $S$ we use an independent
steganalyser $S^*$. We define a filter $F^{(0)}$ that is special for steganalysis
applications (see \cite{deep_stego_best}, \cite{deeplearningstego}, \cite{tan2014stacked}),
as follows
\begin{equation*}
  F^{(0)} = \frac{1}{12}
  	\begin{pmatrix}
      -1 &  2 & - 2 &  2 & -1 \\ 
       2 & -6 &   8 & -6 &  2 \\ 
      -2 &  8 & -12 &  8 & -2 \\ 
       2 & -6 &   8 & -6 &  2 \\ 
      -1 &  2 & - 2 &  2 & -1
    \end{pmatrix} \,. 
\end{equation*}
The structure of the individual steganalyser $S^*$ has the form:
2D convolution with $F^{(0)}$ filter
  $\rightarrow$ Conv2D
  $\rightarrow$ Conv2D
  $\rightarrow$ Max Pooling
  $\rightarrow$ Conv2D
  $\rightarrow$ Conv2D
  $\rightarrow$ Max Pooling
  $\rightarrow$ Fully connected layer ($1024$ neurons)
  $\rightarrow$ Fully connected layer ($1$ neuron)
  $\rightarrow$ Sigmoid function for output.
This structure provides state-of-the-art steganalysis accuracy, \cite{deep_stego_best}, and the
filter $F^{(0)}$ allows to increase convergence speed of the steganalyser $S^*$ training. 
    
For training of this steganalyser we use the Adam optimization algorithm on the loss \eqref{eq:sgan_loss}
with the learning rate equal to $5^{-6}$, $\beta_1 = 0.9$, $\beta_2 = 0.999$. As a loss function
we use a binary cross-entropy.

The setup of experiments can be described as follows:
\begin{itemize}
  \item We train and use the SGAN and/or DCGAN model to generate images to be used as containers;
  \item We train the independent steganalyser $S^*$ using either real images (sec.~\ref{RIS})
  or generated images (sec.~\ref{GIS});
  \item We measure the accuracy of the steganalyser $S^*$.
\end{itemize}

\subsubsection{Training/Testing on Real Images}
\label{RIS}

In this set of experiments we train the independent steganalyser $S^*$ on real images. Results are
provided in tab.~\ref{tab:real_exp}.
\begin{table}[t!]%\footnotesize
	\caption{Accuracy of the steganalyser $S^*$ trained on real images}
	\label{tab:real_exp}
	\begin{center}
		\begin{tabular}{|c|c|c|}
			\hline Type of a test set $\backslash$ Image generator & SGANs & DCGANs\\ 
			\hline Real images & \multicolumn{2}{|c|}{0.962}\\ 
			\hline Generated images &  0.501 & 0.522\\ 
			\hline 
		\end{tabular} 
	\end{center}
\end{table}
From the results we conclude that even the usual DCGAN generate synthetic container images, that can
easily deceive a steganalyser. Although containers generated by on the SGAN model, which is specially
designed to deceive the steganalyser, improve on the result of DCGAN, the improvement is not significant.
These results also suggest that $S^*$ may have been overfit to the data distribution of the real
images, and failed to generalize to the GAN approximation of the data distribution in the synthetic
image domain.

Another interesting result follows from an experiment with HUGO steganography algorithm and SGAN
generated containers:
\begin{itemize}
  \item We use the $\pm1$-embedding algorithm for training the SGAN model;
  \item We test the HUGO steganography analyzer on real images and on images,
  generated by the SGAN model.
\end{itemize}
Containers generated by the SGAN model decrease HUGO steganalysis accuracy from $0.624$ to $0.499$,
which demonstrates that the Steganographic Generative Adversarial Networks can potentially be used
as a universal tool for generating Steganography containers tuned to deceive any specific steganalysis algorithm. 

\subsubsection{Training/Testing on Generated Images}
\label{GIS}

In this set of experiments we train the steganalyser $S^*$ on images, generated using the DCGAN model
from the prior noise distribution $p_{noise}(z)$ for some fixed seed value. In all experiments the size
of the train set is fixed at $\approx 160\,000$. After training we test the analyser $S^*$ on images
generated with the DCGAN model according to the following experimental setups:
\begin{itemize}
  \item[$\mathbf{C1}$.] We use the same seed value;
  \item[$\mathbf{C1}$.] We use the same seed value, as in $\mathbf{C1}$,
  and we additionally tune the DCGAN model for several epochs.
\end{itemize}

\begin{table}[h!tb]%\footnotesize
  \centering
  \caption{Training/testing on generated images according to experimental conditions
  $\mathbf{C1}$-$\mathbf{C3}$}
  \label{tab:diff_seed}
  \begin{tabular}{|c|c|}
    \hline Experimental Conditions & Accuracy \\
    \hline $\mathbf{C1}$ & 0.982 \\
    \hline $\mathbf{C3}$ & 0.499 \\
    \hline
  \end{tabular}
\end{table}

The experiment results in tab.~\ref{tab:diff_seed} indicate that additional tuning of the image generator $G$ make this effect even more pronounced.

We also conduct an experiment with classification of generated images without steganographic
embeddings. For this purposes we train a DCGAN conditional model on the MNIST dataset, and
train a separate classifier for the MNIST classification task. The trained classifier achieved
almost perfect accuracy both on the held-out real MNIST dataset, and on synthetic images
produced by the DCGAN. This provides evidence that it is possible to train an image classifier
that shows acceptable accuracy both on real and synthetic images. However it is the artificial
generation of image containers that breaks the usual approaches to steganalysis.
\section{Information Encryption with SEGAN}
\subsection{Data description}
For the experiments in the current section we use MNIST \cite{lecun1998mnist} and CIFAR-10 \cite{cifar} data sets. Both datasets can be considered as a benchmark in Deep Learning and Compute Vision.

\begin{itemize}
\item The MNIST dataset is a set of gray handwritten digits from 0 to 9 of size $28\times28$;

\item The CIFAR-10 dataset consists of 60000 $32\times32$ RGB images of totally 10 classes: airplane, automobile, bird, cat, deer, dog, frog, horse, ship, truck.		
\end{itemize}

\subsection{Encryption experiments}
We consider SEGAN model as a model for information encryption. It allows to generate images with hidden information inside. In the experimental section we should 
\begin{itemize}
\item check if the generated images are realistic;
\item look at the quality of encryption.
\end{itemize}

\begin{figure}[h!]
	\centering
	\includegraphics[width=9cm]{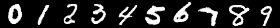}\\
    \includegraphics[width=9cm]{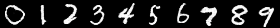}\\
    \includegraphics[width=9cm]{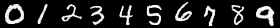}\\
	\caption{Sample synthetic images: MNIST}
	\label{fig:mnist_samples}
\end{figure}

The examples of generate images for MNIST dataset are presented in Fig. \ref{fig:mnist_samples}. We observe the full realism in the generated images.

\begin{figure}[h!]
	\centering
    \includegraphics[width=9cm]{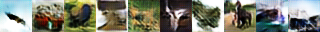}\\
    \includegraphics[width=9cm]{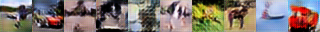}\\
    \includegraphics[width=9cm]{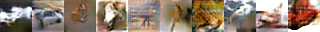}\\
        \includegraphics[width=9cm]{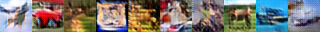}\\
    \includegraphics[width=9cm]{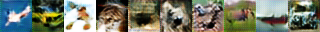}\\
	\caption{Sample synthetic images: CIFAR-10}
	\label{fig:cifar_samples}
\end{figure}

The examples of generate images for CIFAR dataset are presented in  Fig. \ref{fig:cifar_samples}. This images are small and looks quit realistic. This level of realism occurs because of current limitation of generative modeling.

The quality of encryption is presented in Table \ref{tab:segan_qual}. As a measure of quality we considered 

$$
\mathbb{E}_{M}[\#\{i: M_i \neq B(A(M_i))\}],
$$
which means the average percentage of reconstructed pixels. As we can see, our new model allows to encrypt-decrypt messages with the quality of almost 1. It is a little bit harder to encrypt longer messages that short.

\begin{table}[h!]
\centering
\caption{Quality of encryption/decryption (\% of reconstruction bits)}
\label{tab:segan_qual}
\begin{tabular}{|c|c|c|c|}
\hline
Dataset \textbackslash Number of bits & \textbf{16 bit} & \textbf{32 bit} & \textbf{64 bit} \\ \hline
MNIST                                 & 99.98           & 99.54           & 98.65           \\ \hline
CIFAR-10                              & 99.96           & 99.82            &   99.81         \\ \hline
\end{tabular}
\end{table}

\section{Conclusions}
In this work,
\begin{enumerate}
  \item We open a new field for applications of Generative Adversarial Networks, namely,
  container generation for steganography applications;
  \item We consider the $\pm1$-embedding algorithm and test novel approaches to more
  steganalysis-secure information embedding: we demonstrate that both SGAN and DCGAN models are capable of decreasing the detection accuracy of a steganalysis method almost to that of a random classifier;
\item A model for secure adaptive steganographic containers generation has been presented;
\item A number of ways to adopt in order to deceive steganalysis has been proposed;
\item A new GAN-based steganography model has been proposed and tested on MNIST and CIFAR-10 dataset;
% \vspace{0.5cm}
\item As a result an article (\cite{volkhonskiy2017steganographic}) has been written and submitted to NIPS 2016, Workshop on Adversarial Training. This paper cross-checks and significantly extends results of that initial paper.
\end{enumerate}

\bibliographystyle{plain}
\bibliography{main}

\begin{thebibliography}{10}

\bibitem{cheddadetal2010}
Abbas Cheddad, Joan Condell, Kevin Curran, and Paul Mc~Kevitt.
\newblock Digital image steganography: Survey and analysis of current methods.
\newblock {\em Signal Processing}, 90(3):727 -- 752, 2010.

\bibitem{durugkaretal2016}
Ishan Durugkar, Ian Gemp, and Sridhar Mahadevan.
\newblock Generative multi-adversarial networks.
\newblock {\em arXiv preprint arXiv:1611.01673}, 2016.

\bibitem{filleretal2011}
Tom{\'a}{\v{s}} Filler, Jan Judas, and Jessica Fridrich.
\newblock Minimizing additive distortion in steganography using
  syndrome-trellis codes.
\newblock {\em IEEE Transactions on Information Forensics and Security},
  6(3):920--935, Sept 2011.

\bibitem{fridrichkodovsky2012}
Jessica Fridrich and Jan Kodovsk{\`y}.
\newblock Rich models for steganalysis of digital images.
\newblock {\em IEEE Transactions on Information Forensics and Security},
  7(3):868--882, 2012.

\bibitem{ganinetal2015}
Yaroslav {Ganin}, Evgeniya {Ustinova}, Hana {Ajakan}, Pascal {Germain}, Hugo
  {Larochelle}, François {Laviolette}, Mario {Marchand}, and Victor
  {Lempitsky}.
\newblock Domain-adversarial training of neural networks.
\newblock {\em ArXiv e-prints}, May 2015.

\bibitem{goodfellow2016}
Ian {Goodfellow}.
\newblock Nips 2016 tutorial: Generative adversarial networks.
\newblock {\em ArXiv e-prints}, December 2017.

\bibitem{goodfellowetal2014}
Ian Goodfellow, Jean Pouget-Abadie, Mehdi Mirza, Bing Xu, David Warde-Farley,
  Sherjil Ozair, Aaron Courville, and Yoshua Bengio.
\newblock Generative adversarial nets.
\newblock pages 2672--2680, 2014.

\bibitem{huseletal2017}
Martin {Heusel}, Hubert {Ramsauer}, Thomas {Unterthiner}, Bernhard {Nessler},
  G{\"u}nter {Klambauer}, and Sepp {Hochreiter}.
\newblock Gans trained by a two time-scale update rule converge to a nash
  equilibrium.
\newblock {\em ArXiv e-prints}, June.

\bibitem{holubfridrich2010}
Vojt{\v{e}}ch Holub and Jessica Fridrich.
\newblock Designing steganographic distortion using directional filters.
\newblock In {\em WIFS}, 2012.

\bibitem{holubetal2014}
Vojt{\v{e}}ch Holub, Jessica Fridrich, and Tom{\'a}{\v{s}} Denemark.
\newblock Universal distortion function for steganography in an arbitrary
  domain.
\newblock {\em EURASIP Journal on Information Security}, 2014(1):1--13, 2014.

\bibitem{huetal2017}
Zhiting {Hu}, Zichao {Yang}, Ruslan {Salakhutdinov}, and Eric~P. {Xing}.
\newblock On unifying deep generative models.
\newblock {\em ArXiv e-prints}, June 2017.

\bibitem{ioffeszegedy2015}
Sergey Ioffe and Christian Szegedy.
\newblock Batch normalization: Accelerating deep network training by reducing
  internal covariate shift.
\newblock {\em arXiv preprint arXiv:1502.03167}, 2015.

\bibitem{isolaetal2016}
Phillip {Isola}, Jun-Yan {Zhu}, Tinghui {Zhou}, and Alexei~A. {Efros}.
\newblock Image-to-image translation with conditional adversarial networks.
\newblock {\em ArXiv e-prints}, November 2016.

\bibitem{kingmaba2014}
Diederik Kingma and Jimmy Ba.
\newblock Adam: A method for stochastic optimization.
\newblock {\em arXiv preprint arXiv:1412.6980}, 2014.

\bibitem{kingma2014adam}
Diederik Kingma and Jimmy Ba.
\newblock Adam: A method for stochastic optimization.
\newblock {\em arXiv preprint arXiv:1412.6980}, 2014.

\bibitem{kingmaetal2014}
Diederik~P. {Kingma}, Danilo~J. {Rezende}, Shakir {Mohamed}, and Max {Welling}.
\newblock Semi-supervised learning with deep generative models.
\newblock {\em ArXiv e-prints}, June 2014.

\bibitem{kingmawelling2013}
Diederik~P. {Kingma} and Max {Welling}.
\newblock Auto-encoding variational bayes.
\newblock {\em ArXiv e-prints}, December 2013.

\bibitem{cifar}
Alex Krizhevsky and G~Hinton.
\newblock Convolutional deep belief networks on cifar-10.
\newblock {\em Unpublished manuscript}, 40, 2010.

\bibitem{lecun1998mnist}
Yann LeCun.
\newblock The mnist database of handwritten digits.
\newblock {\em http://yann. lecun. com/exdb/mnist/}, 1998.

\bibitem{liu2015faceattributes}
Ziwei Liu, Ping Luo, Xiaogang Wang, and Xiaoou Tang.
\newblock Deep learning face attributes in the wild.
\newblock In {\em Proceedings of International Conference on Computer Vision
  (ICCV)}, 2015.

\bibitem{maitietal2011}
Chandreyee Maiti, Debanjana Baksi, Ipsita Zamider, Pinky Gorai, and
  Dakshina~Ranjan Kisku.
\newblock {\em Data Hiding in Images Using Some Efficient Steganography
  Techniques}, pages 195--203.
\newblock Springer Berlin Heidelberg, Berlin, Heidelberg, 2011.

\bibitem{nowozinetal2016}
Sebastian {Nowozin}, Botond {Cseke}, and Ryota {Tomioka}.
\newblock $f$-gan: Training generative neural samplers using variational
  divergence minimization.
\newblock {\em ArXiv e-prints}, June 2016.

\bibitem{pevnyetal2010a}
Tom{\'a}{\v{s}} Pevn{\`y}, Patrick Bas, and Jessica Fridrich.
\newblock Steganalysis by subtractive pixel adjacency matrix.
\newblock {\em IEEE Transactions on Information Forensics and Security},
  5(2):215--224, 2010.

\bibitem{pevnyetal2010b}
Tom{\'a}{\v{s}} Pevn{\'y}, Tom{\'a}{\v{s}} Filler, and Patrick Bas.
\newblock Using high-dimensional image models to perform highly undetectable
  steganography.
\newblock In {\em Information Hiding}, page 2010, Calgary, Canada, June 2010.

\bibitem{pibreetal2015}
Lionel Pibre, Pasquet {J{\'e}r{\^o}me}, Dino Ienco, and Marc Chaumont.
\newblock Deep learning for steganalysis is better than a rich model with an
  ensemble classifier, and is natively robust to the cover source-mismatch.
\newblock {\em ArXiv e-prints}, November 2015.

\bibitem{deep_stego_best}
Lionel Pibre, Pasquet J{\'e}r{\^o}me, Dino Ienco, and Marc Chaumont.
\newblock Deep learning for steganalysis is better than a rich model with an
  ensemble classifier, and is natively robust to the cover source-mismatch.
\newblock {\em arXiv preprint arXiv:1511.04855}, 2015.

\bibitem{qianetal2015}
Yinlong Qian, Jing Dong, Wei Wang, and Tieniu Tan.
\newblock Deep learning for steganalysis via convolutional neural networks.
\newblock In {\em SPIE/IS\&T Electronic Imaging}, pages 94090J--94090J.
  International Society for Optics and Photonics, 2015.

\bibitem{deeplearningstego}
Yinlong Qian, Jing Dong, Wei Wang, and Tieniu Tan.
\newblock Deep learning for steganalysis via convolutional neural networks.
\newblock In {\em IS\&T/SPIE Electronic Imaging}, pages 94090J--94090J.
  International Society for Optics and Photonics, 2015.

\bibitem{radfrodetal2015}
Alec Radford, Luke Metz, and Soumith Chintala.
\newblock Unsupervised representation learning with deep convolutional
  generative adversarial networks.
\newblock {\em arXiv preprint arXiv:1511.06434}, 2015.

\bibitem{tan2014stacked}
Shunquan Tan and Bin Li.
\newblock Stacked convolutional auto-encoders for steganalysis of digital
  images.
\newblock In {\em Asia-Pacific Signal and Information Processing Association,
  2014 Annual Summit and Conference (APSIPA)}, pages 1--4. IEEE, 2014.

\bibitem{volkhonskiy2017steganographic}
Denis Volkhonskiy, Ivan Nazarov, Boris Borisenko, and Evgeny Burnaev.
\newblock Steganographic generative adversarial networks.
\newblock {\em Workshop on Adversarial Training, Neural Information Processing
  Systems}, 2016.

\end{thebibliography}

\end{document}